\documentclass[preprint,showpacs,preprintnumbers,amsmath,amssymb]{revtex4}

\usepackage{graphicx}
\usepackage{dcolumn}
\usepackage{bm}
\usepackage{color}
\usepackage{mathrsfs}

\newcommand{\llgg}{{\tiny \raisebox{0.0ex}{$<$}\hspace{-0.75em}\raisebox{-1.5ex}{$>$}}}

\begin{document}

\preprint{APS/123-QED}

\title{Electronic Transport in Fullerene C$_{20}$ Bridge Assisted by Molecular Vibrations}

\author{Takahiro Yamamoto$^{1,3}$}
\author{Kazuyuki Watanabe$^{1,3}$}
\author{Satoshi Watanabe$^{2,3}$}
\affiliation{$^{1}$Department of Physics, Tokyo University of Science, 1-3 Kagurazaka, Shinjuku-ku, Tokyo 162-8601, Japan}
\affiliation{$^{2}$Department of Materials Engineering, The University of Tokyo, 7-3-1 Hongo, Bunkyo-ku, Tokyo 113-8656, Japan}
\affiliation{$^{3}$CREST, Japan Science and Technology Agency, 4-1-8 Honcho Kawaguchi, Saitama 332-0012, Japan.}

\date{\today}
\begin{abstract}
The effect of molecular vibrations on electronic transport is investigated with the smallest fullerene C$_{20}$ bridge, utilizing the Keldysh nonequilibrium Green's function techniques combined with the tight-binding molecular-dynamics method. Large discontinuous steps appear in the differential conductance when the applied bias-voltage matches particular vibrational energies. The magnitude of the step is found to vary considerably with the vibrational mode and to depend on the local electronic states besides the strength of electron-vibration coupling. On the basis of this finding, a novel way to control the molecular motion by adjusting the gate voltage is proposed.
\end{abstract}


\maketitle
Inelastic transport associated with local heating in nanoscale devices has been a growing interest in the fields of nanoscience and nanotechnology over the past few years~\cite{rf:Smit,rf:Agrait,rf:Ness,rf:Mont,rf:Chen}. Fullerenes are considered promising candidates for basic elements in nanoscale devices, therefore, electronic transport in fullerene bridges has received significant attention both experimentally~\cite{rf:Park,rf:Pasupathy,rf:Joachim} and theoretically~\cite{rf:Nakanishi,rf:Palacios}. McEuen and co-workers measured current-voltage ($I$-$V$) characteristics of the C$_{60}$ and C$_{140}$ bridges~\cite{rf:Park,rf:Pasupathy} and reported that coupling between electronic and vibrational degrees of freedom plays an important role in electronic transport.

The strength of the electron-vibration coupling is known to be enhanced with decreasing the diameter of the fullerene~\cite{rf:Devos-L}. Thus, among fullerenes, the smallest fullerene C$_{20}$ is expected to have the largest electron-vibration coupling. Recently, the C$_{20}$ fullerene was synthesized by Prinzbach {\it et al}~\cite{rf:Prinzbach}. 
To the best of our knowledge, the issue of inelastic transport through the C$_{20}$ bridge in the presence of electron-vibration scattering has not been addressed. Although elastic transport through the C$_{20}$ bridge has been theoretically investigated by several authors~\cite{rf:Miyamoto,rf:Roland,rf:Otani}, characteristic features remain to be elucidated.

The aim of this Letter is to clarify the role of molecular vibrations in electronic transport through the C$_{20}$ connected to Au electrodes. A semi-infinite one-dimensional (1D) chain is used as a simple and ideal model for the Au electrode, as shown in Fig.~\ref{fig:1}. In an effort to achieve the aim, molecular vibrations in the C$_{20}$ are first examined using the tight-binding molecular-dynamics (TBMD) method~\cite{rf:Wang}. The influence of molecular vibrations on electronic transport characteristics of the C$_{20}$ bridge is discussed in terms of the Keldysh nonequilibrium Green's function (NEGF) method~\cite{rf:Keldysh}, a powerful method used to analyze inelastic transport in the presence of electron-vibration interactions~\cite{rf:Caroli,rf:Datta}. Recently, the NEGF formalism has been successfully applied to inelastic transports in several nanostructures, such as atomic wires~\cite{rf:Frederiksen1,rf:Frederiksen2} and molecular junctions~\cite{rf:Zhu,rf:Galperin,rf:Pecchia,rf:Asai}.

Before studying the vibrational properties of the C$_{20}$ bridge, the stable structure of C$_{20}$ is determined using the TBMD method. The structure of C$_{20}$ is distorted from the highest possible $I_h$ symmetry to the $D_{3d}$ symmetry after the optimization, due to the Jahn-Teller effect. The optimized structure of the C$_{20}$ contains three types of nonequivalent atoms, labeled $a$, $b$, and $c$ in Fig.~\ref{fig:1}. There are four distinct bond lengths: $a_{ab}=$1.464~{\AA}, $a_{bc}=$1.469~{\AA}, $a_{cc'}=$1.519~{\AA}, and $a_{cc''}=$1.435~{\AA}. The obtained bond lengths differ by less than 1~$\%$ from those obtained using the first-principles calculations~\cite{rf:Galli}.

\begin{table}[t]
  \caption{Vibrational energies (meV) of the isolated and connected C$_{20}$.}
  \begin{ruledtabular}
    \begin{center}
      \begin{tabular}{rrrrrrrrrrrr|rrr}
      &\multicolumn{11}{c}{Isolated C$_{20}$}&\multicolumn{3}{c}{Connected C$_{20}$}\\
      $A_{1g}$ & & $A_{2g}$ & & $A_{1u}$ & & $A_{2u}$ & & $E_{g}$ & & $E_{u}$ & & $A_{1g}'$ & & $A_{2u}'$\\
      \hline  
       58.2 & & 66.8 & &  69.2 & &  69.5 & &  29.9 & &  68.9 & &  62.6 & &  11.1\\
       98.6 & & 88.9 & &  74.0 & &  88.6 & &  56.1 & &  72.1 & & 100.8 & &  72.0 \\
      105.2 & &140.2 & & 131.5 & & 110.9 & &  76.8 & &  73.4 & & 106.6 & &  91.4 \\
      135.1 & &      & & 151.3 & & 127.6 & &  94.6 & &  92.4 & & 135.3 & & 112.4 \\
      140.9 & &      & &       & & 148.8 & & 119.9 & & 113.3 & & 141.4 & & 127.7 \\
      161.6 & &      & &       & &       & & 132.6 & & 132.4 & & 161.6 & & 149.2 \\
            & &      & &       & &       & & 137.7 & & 135.9 & &       & &       \\
            & &      & &       & &       & & 148.8 & & 146.6 & &       & &       \\
            & &      & &       & &       & & 163.6 & & 155.7 & &       & &       \\
      \end{tabular}      
    \end{center}
   \label{tab:spectra}
  \end{ruledtabular}
\end{table}

The vibrational energies of the isolated C$_{20}$ are first calculated by diagonalizing a 60$\times$60 force-constant matrix derived from the TBMD calculations. The results of the vibrational energies are classified using the irreducible representations of the $D_{3d}$ symmetry group, as listed in Table~\ref{tab:spectra}. Similarly, the vibrational energies of the C$_{20}$ connected by two springs to the Au electrodes are calculated. The springs are attached to the $a$-atoms on either side of the principal axis of $D_{3d}$ symmetry of the C$_{20}$, as shown in Fig.~\ref{fig:1}. The spring constant was assumed to be $4.37$~eV/{\r A}$^2$, an experimental value for the C$_{60}$ attached to Au electrodes~\cite{rf:Park}. In the present calculations, atomic vibrations of Au atoms were neglected because an Au atom is considerably heavier than a C atom. The connection of the electrodes only influences the vibrational modes with $A_{1g}$ and $A_{2u}$ symmetries, because modes with other symmetries show no stretching of the springs between the C$_{20}$ and Au electrodes. The vibrational energies shifted by the connection are listed in the $A_{1g}'$ and $A_{2u}'$ columns in Table~\ref{tab:spectra}. It is important to note that the vibrational mode of 11.1~meV in the $A_{2u}'$ column corresponds to the shuttle motion of C$_{20}$ that goes back and forth between the two Au electrodes. Using these 55 vibrational modes, the Hamiltonian for the molecular vibrations of the C$_{20}$ bridge can be written as ${\mathscr H}_{\rm vib}=\sum_\lambda\hbar\omega_\lambda(b_\lambda^{\dagger}b_\lambda+\frac{1}{2})$, where $b_\lambda^\dagger$ ($b_\lambda$) is the creation (annihilation) operator of the vibrational quanta ({\it i.e.}, phonon) with energy $\hbar\omega_\lambda$. 

The Hamiltonian for conduction electrons in the C$_{20}$ bridge by the tight-binding model within the H{\" u}ckel approximation is subsequently described. The tight-binding Hamiltonian is expressed by the sum of five parts: ${\mathscr H}_{\rm el}={\mathscr H}_{\rm L}+{\mathscr H}_{\rm LM}+{\mathscr H}_{\rm M}+{\mathscr H}_{\rm MR}+{\mathscr H}_{\rm R}$. ${\mathscr H}_{\rm M}$ represents the Hamiltonian for an extended molecule, Au-C$_{20}$-Au, including the edge Au atoms in the semi-infinite 1D electrodes, ${\mathscr H}_{\rm L/R}$ for the left/right electrodes without the edge atom, and ${\mathscr H}_{\rm LM/MR}$ for the contact between them. The Hamiltonian for the extended molecule is expressed as ${\mathscr H}_{\rm M}=\sum_i\epsilon_{i}c_{i}^{\dagger}c_{i}+\sum_{i,j}t_{ij}^0(c_{i}^{\dagger}c_{j}+{\rm h.c.})$, where $\epsilon_i$ is the on-site energy of $\pi$ orbitals for C atoms and of $6s$ orbitals for Au atoms, $t_{ij}^0$ is a hopping parameter between $i$th and $j$th orbitals in equilibrium, and $c_i^\dagger$ ($c_i$) is the creation (annihilation) operator of an electron on the $i$th orbital. In the present investigation, the on-site energy $\epsilon_{\rm Au}$ for the Au electrode is assumed to lie in the center of the HOMO-LUMO gap of the isolated C$_{20}$. The hopping parameters between Au and C atoms are chosen to be $-1.0$~{eV}, whereas those between $\pi$ orbitals are determined using the TBMD calculations~\cite{rf:Wang}. 

The electron-vibration interaction is given by
\begin{eqnarray}
{\mathscr H}_{\rm e{\text -}v}&=&\sum_{\lambda}\sum_{ij}g_{ij}^\lambda
(c_{i}^{\dagger}c_{j}+{\rm h.c.})(b_\lambda^\dagger+b_\lambda),
\label{eq:el-vib}
\end{eqnarray}
where the coupling constant $g^\lambda_{ij}$ is given as
\begin{eqnarray}
g_{ij}^\lambda=\sum_{\alpha=x,y,z}
\frac{\partial{t_{ij}}}{\partial{s_{i\alpha}}}\sqrt{\frac{\hbar}{2\omega_\lambda}}
\left[\frac{u_{i\alpha}^{\lambda}}{\sqrt{M_i}}-\frac{u_{j\alpha}^{\lambda}}{\sqrt{M_j}}\right],
\end{eqnarray}
where $u_{i\alpha}^{\lambda}$ is an eigenvector of the dynamic matrix and $dt_{ij}/ds_{i\alpha}$ is the hopping modulation with an atomic displacement $s_{i\alpha}$ from equilibrium along the $\alpha$ direction of $i$th carbon atom. The modulations $dt_{ij}/ds_{i\alpha}$ within the C$_{20}$ are determined using the TBMD calculations, and those between the C$_{20}$ and Au electrodes are assumed to be $1.0$~{eV/{\AA}}.

According to the NEGF formalism, the electrical current from the left electrode to a system is given by
\begin{eqnarray}
I=\frac{2e}{h}\int^{\infty}_{-\infty}\!\!\!d\epsilon~{\rm Tr}
\left[{\bm \Sigma}^{<}_{L}(\epsilon){\bm G}^{>}(\epsilon)
-{\bm \Sigma}^{>}_{L}(\epsilon){\bm G}^{<}(\epsilon)\right].
\label{eq:current}
\end{eqnarray}
For the steady-state transport, the lesser and greater Green's functions obey the steady-state Keldysh equation: ${\bm G}^{\llgg}={\bm G}^{r}({\bm \Sigma}^{\llgg}_{\rm L}+{\bm \Sigma}^{\llgg}_{\rm R}+{\bm \Sigma}^{\llgg}_{\rm e{\text -}v}){\bm G}^{a}$, where the self-energy ${\bm \Sigma}^{\llgg}_{\rm L(R)}$ comes from the left (right) contacts and ${\bm \Sigma}^{\llgg}_{\rm e{\text -}v}$ is related to electron-vibration interaction. The retarded (advanced) Green's function is given by the Dyson equation: ${\bm G}^{r(a)}={\bm G}^{0,r(a)}+{\bm G}^{0,r(a)}{\bm \Sigma}^{r(a)}_{\rm e{\text -}v}{\bm G}^{r(a)}$, where ${\bm G}^{0,r(a)}$ is an unperturbed retarded (advanced) Green's function and ${\bm \Sigma}^{r(a)}_{\rm e{\text -}v}$ is the retarded (advanced) self-energy due to electron-vibration interaction. 
In the present calculation, the self-energies due to electron-vibration interaction are treated perturbatively using the Feynman diagram technique, and are expanded up to the lowest order of self-energy diagrams. On the other hand, the self-energies of semi-infinite electrodes can be calculated analytically in the case of the semi-infinite 1D electrode~\cite{rf:Datta}. 
The equation~(\ref{eq:current}) is reduced to the well-known Landauer formula for the elastic current $I_{\rm el}$ by setting ${\bm\Sigma}_{\rm e{\text -}v}^{<}=0$ in the Keldysh equation.

The $I$-$V$ characteristics of the C$_{20}$ bridges are shown in the inset in Fig.~\ref{fig:2}, where the solid and dashed curves represent the total current $I$ and the elastic current $I_{\rm el}$, respectively. From the calculated $I$-$V$ characteristics, the total current curve is found to deviate slightly upward from the elastic curve as the applied bias-voltage increases. 
The dissipation power into molecular vibrations is also estimated, up to $V_{\rm bias}=100$ mV, as of the order of $1$~nW, which is less than $3\%$ of the total power generated in the entire bridge. This means that local heating due to the molecular vibrations is not a serious bottleneck for transport characteristic of the C$_{20}$ bridge compared with that due to contact resistance.

The contribution from each vibrational mode to the transport characteristics can be seen clearly in the differential conductance, $dI/dV_{\rm bias}$, shown in Fig.~\ref{fig:2}. Solid and dashed curves in Fig.~\ref{fig:2} represent the total differential conductance $dI/dV_{\rm bias}$ and its elastic part $dI_{\rm el}/dV_{\rm bias}$, respectively. In contrast to the smooth curve for the elastic differential conductance, large discontinuous jumps appear in the total differential conductance curve at particular bias-voltages, as indicated by arrows in Fig.~\ref{fig:2}. The contribution from vibrational modes other than those indicated by the arrows is increasingly small, despite the nonzero coupling constants $g_{ij}^\lambda$. A typical example is illustrated by the conductance step at $11.1$~mV, originating from the shuttle motion of C$_{20}$, which is two orders of magnitude less than the step at $29.9$~mV. The shuttle motion will be discussed in further detail later.

The physical origin for the considerable variability of the magnitude of the discontinuous step in Fig.~\ref{fig:2} with the vibrational mode, is now discussed from the viewpoint of the electronic structures of the C$_{20}$ bridge. 
Fermi's golden rule gives us not only the scattering rate of electronic states, but also the magnitude of these steps in the conductance curve. According to Fermi's golden rule, the magnitude of the steps is proportional to $|\sum_{i,j}g_{ij}^{\lambda}\Psi_i^*(\epsilon-\hbar\omega_\lambda)\Psi_j(\epsilon)|^2$ . Here $\Psi_i(\epsilon)$ represents a scattering electronic state at the $i$th atom, in the absence of an electron-vibration interaction. The scattering electronic states within the narrow bias-window $[\epsilon_F-eV/2, \epsilon_F+eV/2]$ can be approximately replaced by $\Psi_i(\epsilon_F)$ at the Fermi level, therefore, the magnitude of the conductance steps can be estimated by $S_\lambda\equiv\hbar^2|\sum_{i,j}g_{ij}^{\lambda}G_{ij}^{0,<}(\epsilon_F)|^2$, referred to as the scattering intensity hereafter, where $G_{ij}^{0,<}(\epsilon_F)=i\hbar^{-1}\Psi_i^*(\epsilon_F)\Psi_j(\epsilon_F)$ is the unperturbed lesser Green's function. Figure~\ref{fig:3} shows the scattering intensity $S_\lambda$ for the C$_{20}$ bridge. The intensity exhibits large peaks for particular modes indicated by the arrows. These mode energies coincide completely with the positions of the discontinuous steps indicated by the arrows in Fig~\ref{fig:2}. 
As seen in the expression of $S_\lambda$, the magnitude of the conductance steps is determined by the local electronic states near the Fermi level, as well as vibrational states of the C$_{20}$ bridge. $G_{ij}^{0,<}(\epsilon_F)$ describes the correlation between electrons with the Fermi energy on the $i$th and $j$th atoms and the $g_{ij}^\lambda$ has a large value for strong stretching of interatomic bond, therefore, large conductance-steps are thought to appear when the bond between atoms with high electron-densities stretch strongly.

Thus, a novel way to control the motion of C$_{20}$ between two electrodes under current flow, utilizing the obtained knowledge from the correlation between scattering electronic states and molecular vibrations is proposed. The shuttle motion of C$_{20}$ is a key motion, however, its excitation rate is very low owing to the small $S_\lambda$ at $\hbar\omega_\lambda=11.1$~meV in Fig.~\ref{fig:3}.
As explained previously, the rate is expected to increase significantly when the electronic state localized at C atoms adjacent to Au electrodes lie close to the Fermi level, because only two springs between C$_{20}$ and Au electrodes stretch in the shuttle motion. 
However, such a localized state lies $1.5$~eV below the Fermi level ($\epsilon_F=0$~eV), as shown in Fig.~\ref{fig:4}(a), and the scattering intensity of the shuttle motion exhibits a maximum peak at $-1.5$~eV in Fig.~\ref{fig:4}(c). Therefore, its excitation rate can be enhanced by tuning the gate voltage to shift the localized state to the Fermi level. Of course, more precise analyses beyond the rigid-band picture assumed here will be necessary to quantify the gate-voltage effect on the Fermi level.
The control mechanism of the shuttle motion of C$_{20}$ proposed herein is complementary with the Coulomb blockade mechanism for the recent experiment on electronic transports in the C$_{60}$ {\it weakly} suspended between Au electrodes~\cite{rf:Park}. The mechanism proposed in the current investigation is efficient, even for inelastic electronic transport through fullerenes {\it strongly} connected to the electrodes.

We finally emphasize the extention of the proposed idea to other molecular bridges. For the 1,4-benzenedithiol (BDT=C$_6$H$_4$S$_2$) bridge, it has been theoretically predicted that a rotational motion of 1,4-BDT molecule is strongly coupled to scattering electronic states at low-bias voltages~\cite{rf:Sergueev}. By analogy with our results, the rotational motion will be switched on when a scattering state with high electron-densities on sulfur atoms in 1,4-BDT adjacent to the electrodes is shifted to the Fermi level by tuning the gate voltage.

In conclusion, the inelastic electronic transport assisted by molecular vibrations in the smallest fullerene C$_{20}$ bridge was investigated using the NEGF formalism combined with the TBMD method. The effect of the molecular vibrations on the transport characteristics of C$_{20}$ appear as discontinuous steps with various heights in the differential conductance curve. The stepwise behavior of the conductance, and the variation of magnitude of the step was clearly understood by analyzing the electronic structures based on Fermi's golden rule, as well as the vibrational structures. Local heating caused by the molecular vibrations of the C$_{20}$ bridge were also found not to be a serious bottleneck for functions of fullerene-based nanodevices. Moreover, the idea that the shuttle motion of the fullerenes can be controlled by adequately tuning the gate voltage was proposed. The present study provides a clue toward a more complete understanding of inelastic electronic transport and the electromechanics in various molecular devices.

This work was supported in part by ``Academic Frontier" Project of MEXT (2005-2010). 
Part of the numerical calculations were performed on the Hitachi SR8000s at ISSP, The University of Tokyo.

\newpage
\begin{figure}[t]
  \caption{Schematic diagram of the C$_{20}$ fullerene connected to Au electrodes by two springs. 
  The C$_{20}$ with $D_{3d}$ point-group symmetry contains four distinct bond lengths among 
  nonequivalent carbon atoms labeled $a$, $b$, and $c$ (or $c'$, $c''$).}
  \label{fig:1}
  \caption{The differential conductance in the unit of $2e^2/h$. Solid and dashed curves represent 
  the total differential conductance and its elastic component, respectively. The arrows indicate 
  particular discontinuous steps. The inset represents the total and elastic electronic current 
  through the C$_{20}$ bridge.}
  \label{fig:2}
  \caption{The scattering intensity of electrons due to molecular vibrations of the C$_{20}$ 
  bridge. The arrows indicate the particular peaks that coincide with the discontinuous steps 
  in Fig.~\ref{fig:2}.}
  \label{fig:3}
  \caption{The electron density of Au-C$_{20}$-Au at (a) $-1.5$~eV and (b) $0$~eV (the Fermi level) 
  and (c) the scattering intensity for the shuttle motion of C$_{20}$. The electron density is 
  indicated by the shading on the atom spheres.}
  \label{fig:4}
\end{figure}

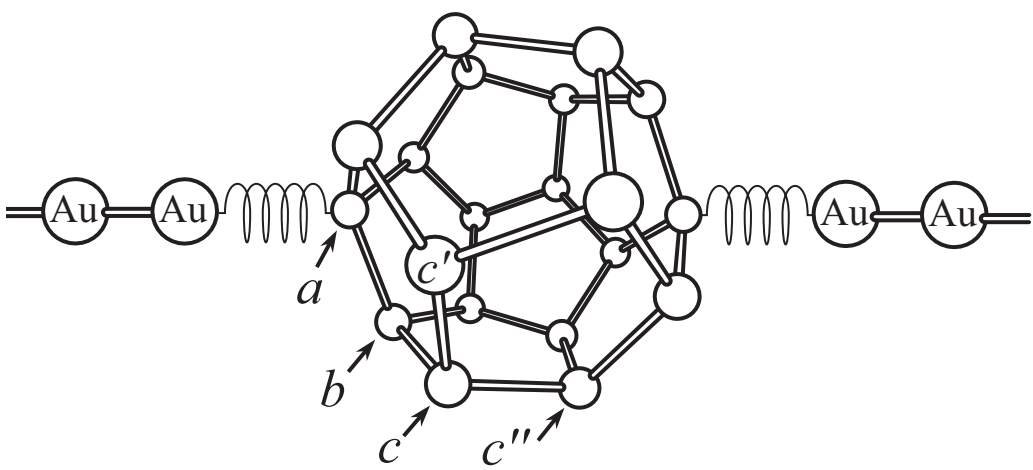
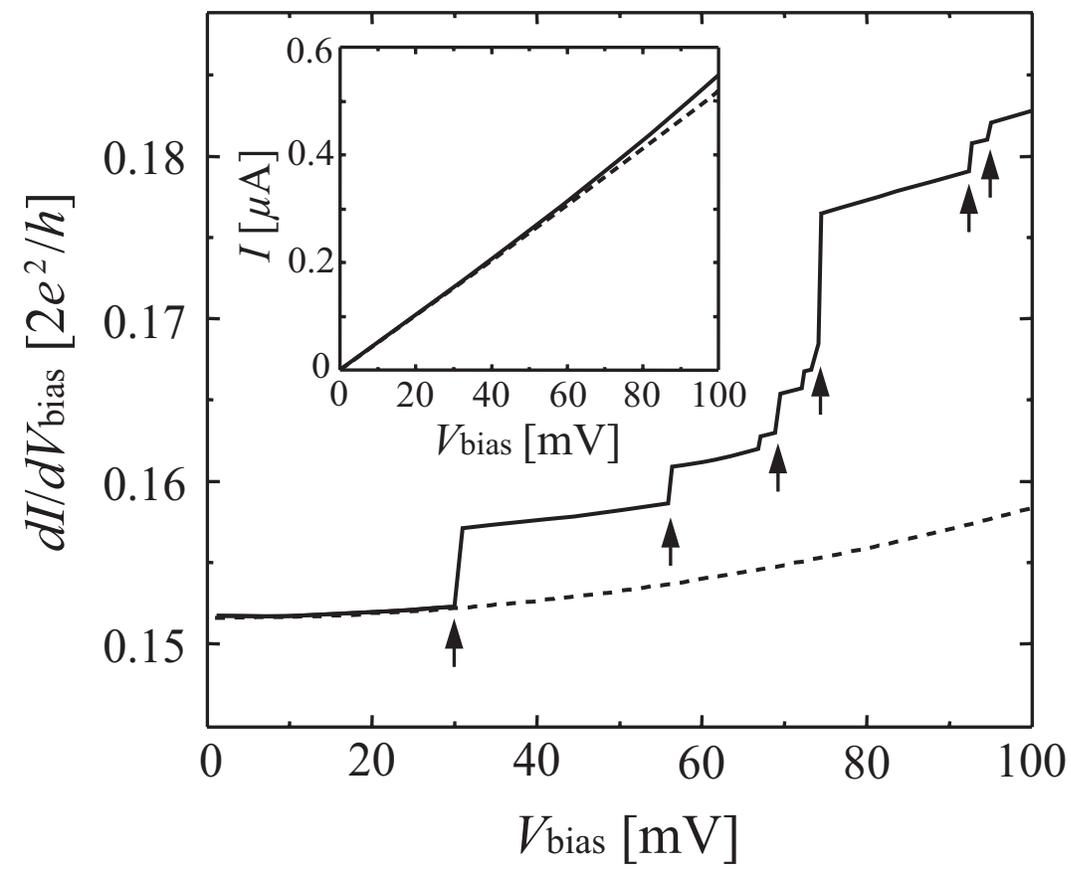
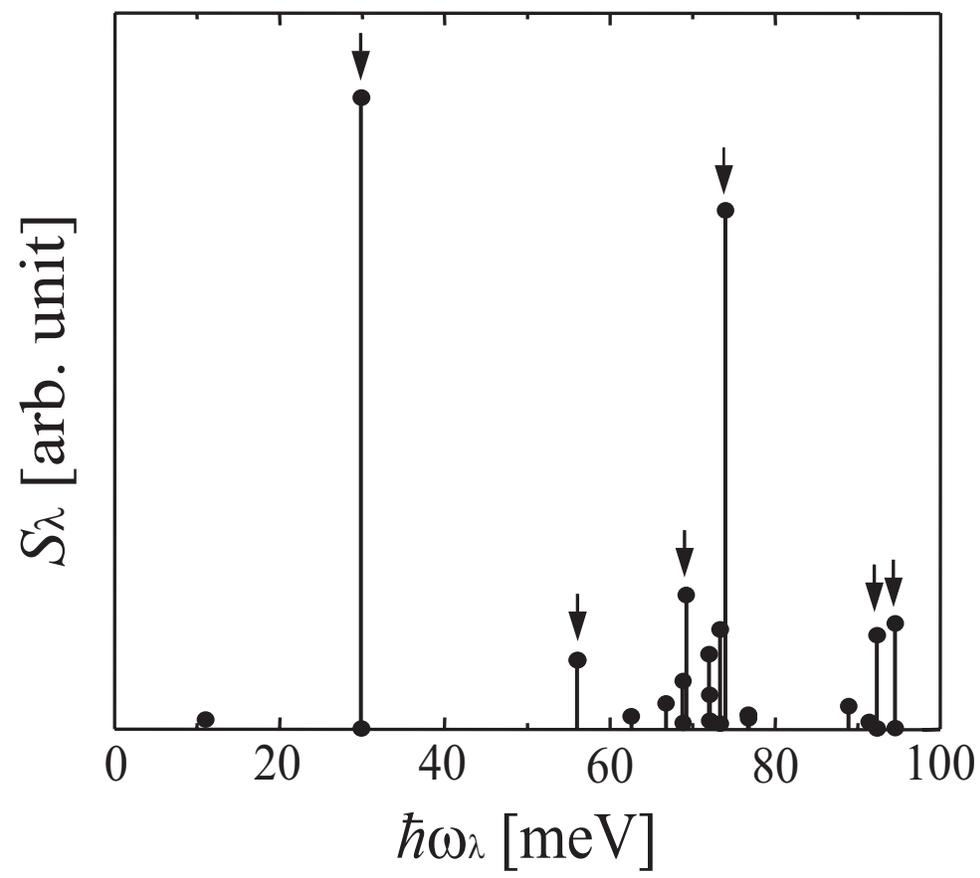
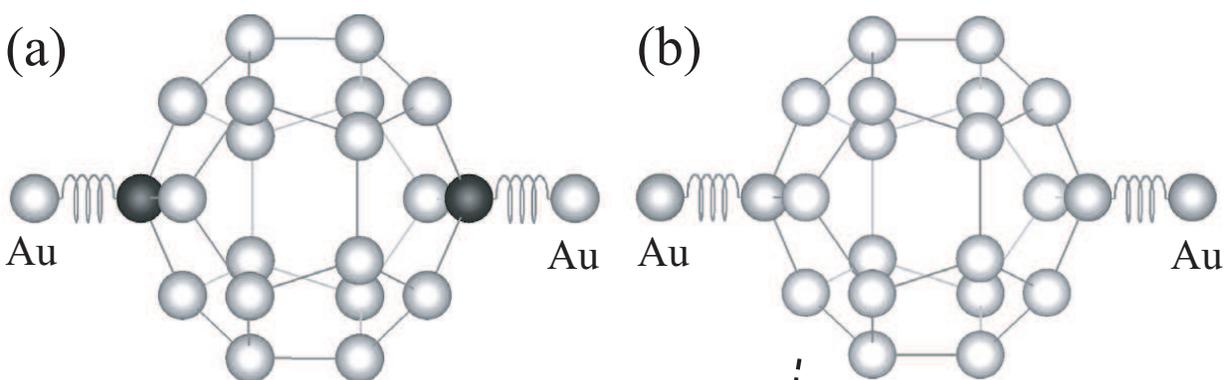
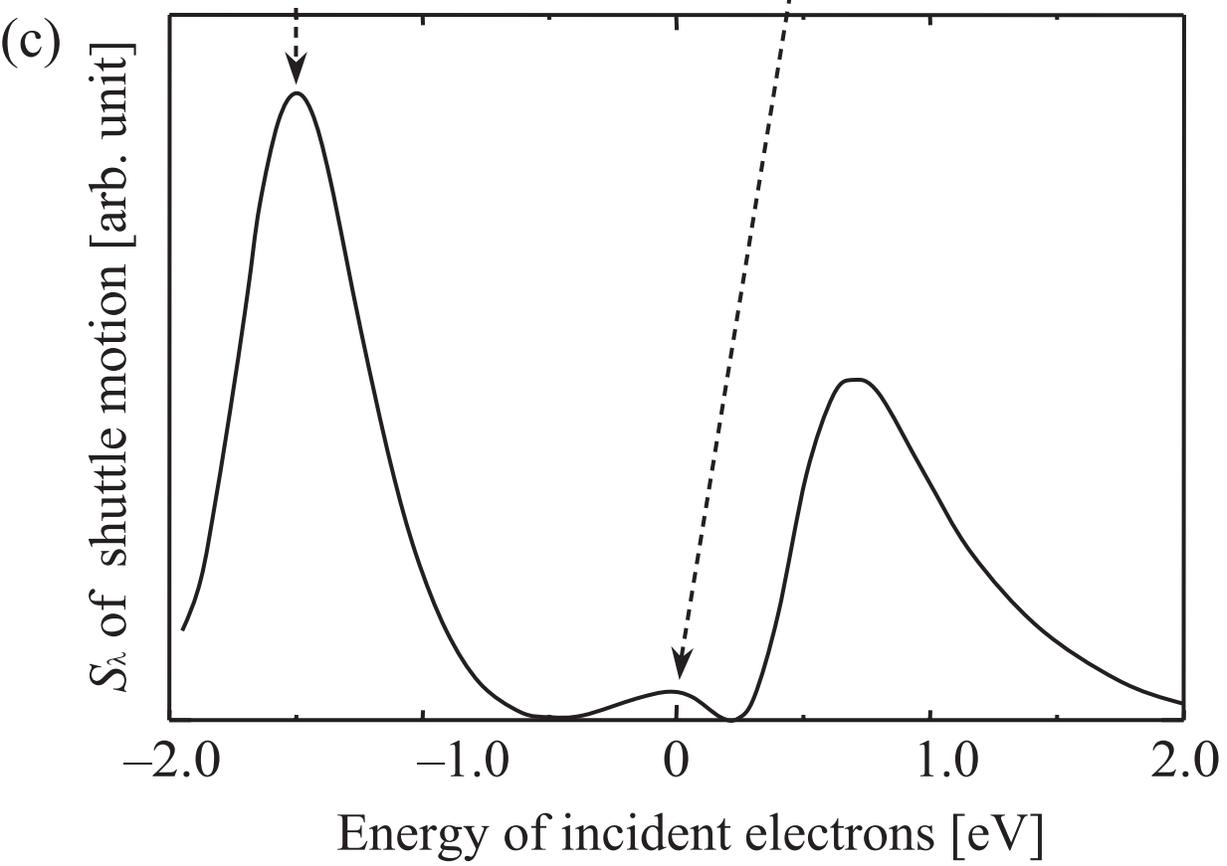
\end{document}